\documentclass[useAMS,usenatbib]{mn2e}

\usepackage{graphicx}\usepackage{epsfig}\usepackage{epsf}
\usepackage{amsmath}\usepackage{amssymb}\usepackage{stfloats}
\title[LBVs and BBS]{The isolation of Luminous Blue Variables resembles
  aging B-type supergiants, not the most massive unevolved stars}

\author[Smith]{Nathan Smith\thanks{E-mail:
    nathans@as.arizona.edu} \\
  Steward Observatory, University of Arizona, 933 N. Cherry
  Ave., Tucson, AZ 85721, USA}

\begin{document}

\pagerange{\pageref{firstpage}--\pageref{lastpage}} \pubyear{2012}
\maketitle
\label{firstpage}

\begin{abstract}

  Luminous blue variables (LBVs) are suprisingly isolated from the
  massive O-type stars that are their putative progenitors in
  single-star evolution, implicating LBVs as binary evolution
  products.  Aadland et al. (A19) found that LBVs are, however, only
  marginally more dispersed than a photometrically selected sample of
  bright blue stars (BBS) in the Large Magellanic Cloud (LMC), leading
  them to suggest that LBV environments may not exclude a single-star
  origin.  In both comparisons, LBVs have the same median separation,
  confirming that any incompleteness in the O-star sample does not
  fabricate LBV isolation.  Instead, the relative difference arises
  because the photometric BBS sample is far more dispersed than known
  O-type stars.  Evidence suggests that the large BBS separation
  arises because it traces less massive ($\sim$20~$M_{\odot}$), aging
  blue supergiants. Although photometric criteria used by A19 aimed to
  select only the most massive unevolved stars, visual-wavelength
  color selection cannot avoid contamination because O and early B
  stars have almost the same intrinsic color.  Spectral types confirm
  that the BBS sample contains many B supergiants.  Moreover, the
  observed BBS separation distribution matches that of
  spectroscopically confirmed early B supergiants, not O-type stars,
  and matches predictions for a $\sim$10~Myr population, not a 3-4 Myr
  population.  A broader implication for ages of stellar populations
  is that bright blue stars are not a good tracer of the youngest
  massive O-type stars.  Bright blue stars in nearby galaxies (and
  unresolved blue light in distant galaxies) generally trace evolved
  blue supergiants akin to SN~1987A's progenitor.
  
\end{abstract}  


\section{INTRODUCTION}

The massive eruptive stars known as luminous blue variables (LBVs) are
critical for understanding the evolution and fates of massive stars.
This is because LBVs have the highest observed mass-loss rates of any
class of stars, and because this mass loss (which may or may not
remove the H envelope) profoundly influences the fate of the star and
the type of eventual supernova (SN) explosion \citep[see][]{smith14}.
Understanding the physical mechanism of this mass loss and its
metallicity dependence is therefore critical for models of stellar
evolution, whether it is driven by normal winds, eruptive events when
a massive star exceeds the Eddington limit, or binary interaction
episodes
\citep{ppod10,so06,groh13,groh13b,justham14,blagovest16,gotberg17}.

The standard view of LBVs has been that they correspond to a very
brief {\it transitional} phase of the most massive single stars, when
the star moves from core H burning to core He burning.  In this view,
LBV winds or eruptions are the prime agent that removes the H envelope
to produce Wolf-Rayet (WR) stars
\citep{langer94,heger03,mm03,meynet11}. This transition from single
O-type stars to WR through their own mass loss is often referred to as
the ``Conti scenario'' \citep{conti76}.  The reliance upon the LBV
phase for making WR stars from single stars is even more accute
because of lowered O-star wind mass-loss rates
\citep{fullerton06,bouret05,so06,smith14}.  It is therefore critical
to have model-independent tests of this single-star evolutionary
paradigm.

Single vs. binary scenarios can be addressed by studying the ages and
environments of LBVs.  For stars at the same place on the HR Diagram,
the age of the surrounding environment can differentiate binary
evolution products from single stars, since mass gainers and mergers
may have had significantly lower initial masses and longer lifetimes
than effectively single supergiant stars of the same current
luminosity.  A clear prediction is that in the single-star scenario,
where LBVs occur immediately after core H exhaustion in transition to
their He burning phase as WR stars, the spatial locations of LBVs
should follow those of massive, young, early O-type stars that are
their immediate progenitors.  At these high initial masses, the
lifetimes are very short (3-4 Myr), and there is not enough time to
move far from their birth sites.  In a binary scenario, on the other
hand, LBVs should be more dispersed than young O-type stars because
they have been rejuvenated after a delay due to their longer
main-sequence lifetime (or they may have received a kick from a
companion's SN), whereas the most massive O-type stars have already
died.

Most stellar age indicators are too imprecise for this task, because
one is interested in being able to distinguish between ages of around
3-4 Myr (main sequence lifetimes of $M_{\rm ZAMS} > 40 M_{\odot}$
stars, for example, appropriate to classical LBVs) or a factor of only
about 2-3 older corresponding to $\sim$20~$M_{\odot}$ stars with
longer lifetimes that have been rejuvented though mass accretion or
mergers.  For example, in the star formation history study of the LMC
by \citet{hz09}, there is one single age bin for all ages $<$9 Myr; so
whether they are single or binary, almost all the LBVs should be
lumped into one bin.  Since LBVs are generally not in clusters, age
estimates based on turnoffs, RSG luminosity, or luminosity functions
\citep{bds19,eldridge17,schneider14} generally can't be applied to
LBVs.  The most reliable clock for the highest mass stars turns out to
be using a spatial association with other stars that must have very
short lifetimes: i.e. early O-type stars.  A spectrum of a single O
star doesn't provide an age, of course, but the relative degree of
clustering of those O-type stars does give a relative statistical age,
because O stars are born in clusters that disperse with time.  These
are the same stars that should be the single-star progenitors of
classical LBVs.  \citet{st15} performed this spatial comparison,
examing the cumulative distributions of separations to the nearest
O-type stars on the sky for O star subtypes, LBVs, WR stars, and other
classes of evolved stars in the Large Magellanic Cloud (LMC).
\citet{st15} found LBVs and LBV candidates to be remarkably isolated
from massive O-type stars, much more so than allowed by single-star
models, thus apparently ruling out the single-star evolutionary
scenario for LBVs.  LBVs in the Milky Way showed a similar avoidance
of O stars, although extinction in the Galactic plane and uncertain
distances made this harder to quantify than in the LMC.  In brief,
LBVs showed a clear preference to avoid massive, young clusters of
O-type stars.  This led \citet{st15} to suggest an alternative
hypothesis that the observed isolation of LBVs could only be
understood if they are primarily the products of close binary
evolution.  In this alternative view, LBVs are not the most massive
single stars in transition, but instead, LBVs are evolved massive blue
straggler stars.

Using a model for the passive dispersal of aging massive stars in
clusters that drift apart with age due to their birth velocity
dispersion, \citet{mojgan17} demonstrated that such a model could
quantitatively explain the observed distribution of O-star subtypes
(early, mid, and late O-type stars).  However, they confirmed that the
same dispersal of clusters could not account for the locations of LBVs
as single stars with ages and initial masses appropriate to their
current luminosity.  LBVs require either much faster drift speeds than
O stars (i.e. kicks from a companion's SN) or older ages commensurate
with those of stars at lower initial masses around 20 $M_{\odot}$.
Interestingly, this blue-straggler view of LBVs as mass gainers or
mergers in binaries also agreed with independent theoretical studies
seeking to understand how LBVs might be SN progenitor stars
\citep{justham14}.

This new blue straggler view of LBVs is in direct contradiction to the
traditional view for their role in stellar evolution.  In addition to
giving a different origin for LBVs themselves, it also has the
consequence of removing LBVs from the single-star evolutionary
scenario, wherein they play a crucial role in removing the H envelope
to make WR stars.  This modification has sparked some debate.  In
particular, \citet{h16} had a different take on subdividing the data,
and preferred the traditional single-star view.  \citet{h16} noticed
that if one excludes most of the LBV sample, then the three most
luminous LBVs in the LMC do have a median separation similar to that
of O-type stars, which in their interpretation supported the
single-star scenario after all. \citet{h16} also pointed out that the
lower-luminosity LBVs have a separation distribution similar to red
supergiants (RSGs), taken as support for a single-star view wherein
these LBVs are post-RSGs from initially 30-40 $M_{\odot}$ stars.  For
both points, however, \citet{smith16} showed that this was a
mischaracterization of the data.  The the most luminous LBVs should
have initial masses of around 50-100 $M_{\odot}$, but the common
O-type stars with a similar spatial distribution noted by \citet{h16}
were dominated by late O-type stars with initial masses around 18-25
$M_{\odot}$.  Similarly, the population of RSGs was dominated by
relatively low initial masses of $\sim$15 $M_{\odot}$, so their
similararity to the low-luminosity LBV distribution (expected to have
single-star initial masses of 30-40 $M_{\odot}$) contradicts a
single-star scenario.  Moreover, \citet{smith16} demonstrated that
there is no significant difference between LBVs and LBV candidates, so
that including ``candidate'' LBVs would not skew the results as
\citet{h16} argued.  (Note that ``candidate'' LBVs are stars with
similar spectra and luminosities to LBVs, often with circumstellar
shells that indicate a prior outburst, but which have not yet been
observed photometrically to undergo LBV eruptions.)

Motivated to weigh in on this debate, Aadland, Massey, Neugent, \&
Drout (2019; A19 hereafter) aimed to provide an independent check on
the isolation of LBVs.  A19 were concerned primarily about how the
unknown level of incompleteness of the spectroscopically confirmed O
star reference sample might skew the results (i.e.  O stars missing
from the sample because they don't have spectra might make LBVs appear
artificially isolated from their nearest known O star neighbors). A19
therefore chose a complimentary approach with different selection
criteria.  Instead of spectroscopically confirmed O-type stars as a
reference for a clustered young massive population, they chose to
compare LBVs to a photometrically selected sample of bright blue stars
(BBS). Their intent was that photometric selection could yield a
complete sample of the most massive unevolved stars in the LMC. Using
the BBS sample as a reference, A19 found the median BBS separation to
be only about 30\% smaller than the LBV median, whereas the median
separation for spectroscopically confirmed O-type stars was 10 times
smaller than for LBVs.  A19 attributed this difference to
incompleteness in the spectrocopically confirmed O stars, and
interpreted the smaller difference from LBVs as not contradicting the
standard picture of massive single-star evolution.

In this paper, we take a closer look at the BBS sample and the
conclusions of A19.  First, in section 2, we point out that the median
separation of LBVs from either BBS or O-type stars was identical in
the two studies of A19 and \citet{st15}, confirming earlier
suggestions \citep{st15} that any incompleteness of the O star sample
has no impact on the apparent isolation of LBVs.  Then we investigate
potential concerns with the BBS sample of A19 and its interpretation,
quantifying the effects of choosing to exclude all the massive O stars
in 30 Doradus (section 3), and quantifying how reliably the color cuts
can select the most massive unevolved stars (section 4), as required
for this comparison.  After demonstrating that color selection cannot
reliably select only the most massive unevolved stars because of
contamination from older B supergiants, we demonstrate (section 5)
that in fact, the distribution of separations for the photometric BBS
sample is practically identical to the spatial distribution of known,
spectroscopically confirmed early B supergiants.  We also comment
(section 6) on the related implications for observed separation
distributions of WR and specifically WN3/O3 stars, which have been
tested with the same methods.  We conclude that the less severe
isolation of LBVs when compared to the photometric BBS sample arises
because the BBS sample is old, not because LBVs are young.

\section{INCOMPLETENESS OF COMPARISON SAMPLES HAS LITTLE IMPACT}

Finding that LBVs are isolated from massive O-type stars overturns a
long-held paradigm of massive star evolution, but it is a statistical
result that could have potential selection bias, and so independent
checks with alternative selection criteria could be valuable.  The
main motivation for undertaking an independent study using a
photometric sample was that A19 were concerned about the possible
incompleteness of spectroscopically confirmed O-type stars, because
not all massive stars in the LMC have known spectral types.  If, for
example, past efforts to gather spectra for massive stars have
concentrated on clustered regions in the LMC, and have therefore
neglected field stars, then there may be additional unknown O-type
stars in the field that are not being counted in the analysis of
spatial separations between LBVs and the nearest O-type star.  A19
were concerned that this incompleteness might skew the results and
cause LBVs to appear artificially isolated.

This potential concern was noted originally by \citet{st15}, who
argued that it wouldn't matter much.  O-type stars are known to reside
mostly in clusters and they essentially provide a map of the space
density of young massive stars.  Adding some O-stars in the field may
serve to raise the quantitative value of the local minimum slightly,
in terms of the number of O stars per unit sky area, and it can
therefore alter the numerical value of the age one infers based on
that space density \citep{mojgan17}.  It does not, however, alter the
fact that O stars have a high concentration in clusters.  Having a
complete count of all the field O-type stars is not needed for this
study.  What is very important is that most of the O star {\it
  clusters} are known, and that LBVs are not in those clusters.  What
would be needed to make LBVs consistent with a single-star scenario
would be to have unrecognized clusters of O stars surrounding each
LBV, which is unlikely given that most LBVs in the LMC have been
imaged with the {\it Hubble Space Telescope} to look for shell
nebulae.

In the end, the results of the analysis conducted by A19 confirmed
that possible incompleteness of the O star sample has no impact on the
outcome.  This is evident from the resulting median of the
distribution of separations between LBVs and the nearest BBS star or O
star.\footnote{Note that the ``nearest'' neigbor excludes possible
  unresolved companions in a binary; both studies refer to the nearest
  spatially resolved stars.}  Using BBS stars as a reference, A19
measured a median separation between LBVs and their nearest BBS
neighbor of 181{\arcsec}.  Using a sample of spectroscopically
confirmed O-type stars as a reference, \citet{st15} measured a median
separation between LBVs and their nearest O-star neighbor of
0.05$^{\circ}$ or 180{\arcsec}.  The results are essentially
identical.  {\it If incompleteness of the spectroscopically confirmed
  O-star sample were to blame for the apparent isolation of LBVs, then
  LBVs would show a smaller median separation when using a ``more
  complete'' sample of young massive stars.}

What happened instead is that the reference sample of BBS stars
shifted to much larger median separation than known O-type stars,
making them more isolated than O stars and therefore more similar to
LBVs.  For their BBS sample, A19 quote a median separation to the
nearest other BBS star of 129{\arcsec} (or a projected separation of
31 pc).  By contrast, known early O-type stars have a median
separation 10 times less, or only 3 pc \citep{st15}.  Mid and
late-type O stars have somewhat larger median separations than early
O-types, but still less than 10 pc \citep{st15}.

At this point, one must question the BBS sample as a tracer of the
most massive unevolved stars, simply because they are not tracing a
clustered population.  The median separation between BBS stars and
their nearest BBS neighbor is 31 pc, and critically, less than about
4\% of the BBS sample has a separation to the nearest neighbor that is
closer than $\sim$5 pc.  By contrast, 70\% of the spectroscopically
selected early O stars have a separation less than 5 pc.  If it were
true that the BBS stars are {\it a complete sample of the most massive
  unevolved single stars}, then this observed distribution would upend
most of what we understand about the birth environments of massive
stars and massive star formation.

It is well established that most O stars are found in clusters and
associations \citep{blaauw64,lynds80,garmany82,gies87}.  From a fairly
complete magnitude-limited sample of bright Galactic O-type stars,
\citet{gies87} estimates that at least 70\% reside in known young
clusters and associations, while the remainder was thought to be a mix
of runaway stars ejected from clusters and some stars that are the
most massive star in a less massive cluster \citep[see
  also][]{renzo19,eldridge11}.  This seems to be in very good
agreement with the observed separation distribution of
spectroscopically confirmed early O-type stars in the LMC
\citep{st15}, but the separation of BBS stars (A19) seems incompatible
with known clustered environments of O-type stars.  The inescapable
conclusion seems to be that the BBS sample must be contaminated by an
older population of evolved bright blue stars in the field.

Understanding why the BBS sample is more dispersed than young O stars
is critical for correctly interpreting the different results found by
A19 and \citet{st15}.  This discussion follows in the next few
sections.

Before that, however, an important point should be made concerning the
mechanics of this sort of comparison.  The analysis method in these
two studies used the observed distributions of separation to a nearest
massive star neighbor as a way to infer {\it relative} ages of
populations of stars, in order to discriminate between single and
binary star evolutionary scenarios.  There are two essential
requirements that must be met for this method to be valid:

First, the reference sample to which populations of stars are being
compared must, in fact, be confidently known to be young.  The way
that the comparison works is that a separation distribution indicates
whether a sample of target stars (in this case LBVs) is as old or
older than a reference sample (in this case, the photometric BBS
sample or O-type stars).  More to the point, the age of that reference
population must be known at least as precisely as the difference in
age one is trying to test for.  Spectroscopy allows one to select a
reference sample of early O-type stars that are certain to be both
massive and young.  While wide-field photometric samples can be useful
to flag issues related to severe incompleteness, it is much more
difficult to guard against contamination from older stars in a
photometric sample, as discussed below. This means that the typical
age of a star in a photometric sample of blue stars is much harder to
judge.  Contamination by older stars will skew a distribution to
larger separations on the sky.

Second, the reference sample to which populations of stars are being
compared must, in fact, be clustered, otherwise the relative spatial
distribution on the sky is not meaningful.  In other words, this test
equates a high degree of clustering with youth.  It relies upon the
assumption that massive stars are mostly born in clusters, and that as
a population of stars ages, they drift apart and the O-type stars die
off, such that the average separation to the nearest O-type star grows
with time.  If the comparison sample is not highly clustered, then
this logic dissolves. The gradual dispersal of clusters accompanied by
removal of the most massive stars as they die was modeled
quantitatively by \citet{mojgan17}, who calculated values for the
expected median separation and separation distributions of such
samples.  In these models, the most massive unevolved stars should
have a typical separation from the nearest other O star of only a few
pc, which again, is found to be in quite good agreement with the
observations of spectroscopically confirmed early O-type stars.  Thus,
whatever the incompleteness may be, the spectroscopic O star sample
behaves as expected and is not strongly affected by incompleteness in
terms of its overall spatial distribition.  On the other hand,
\citet{mojgan17} calculate that a median separation of $\sim$30 pc
corresponds to post-main-sequence ages of around 10 Myr and initial
masses of $\sim$20 $M_{\odot}$. According to the observed median
separation of the BBS sample of 31 pc, one would conjecture that the
BBS sample is dominated by evolved $\sim$20 $M_{\odot}$ stars on
average, not the most massive unevolved stars of 40-100 $M_{\odot}$.
This contamination, rather than single-star evolution, explains A19's
result.  Possible causes of contamination or bias are explored below.

\begin{figure*}
\includegraphics[width=6.0in]{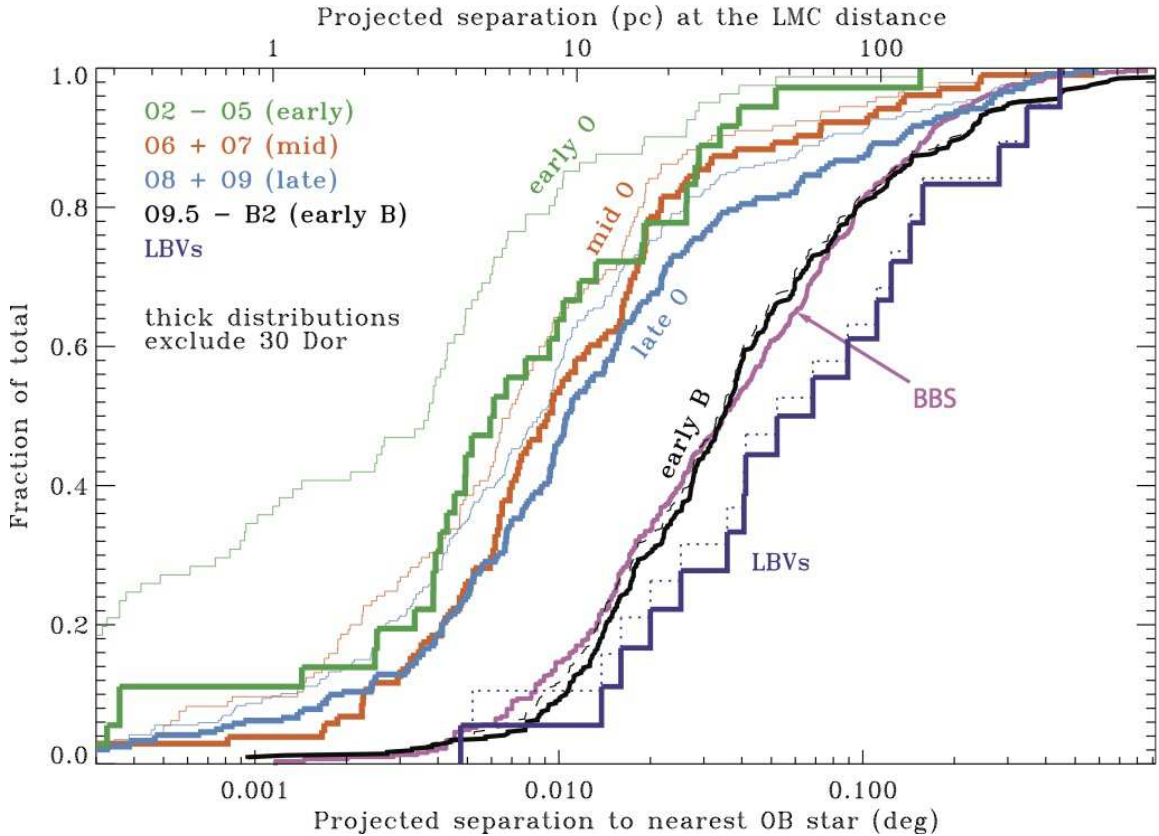}
\caption{Cumulative distribution of separations from the nearest O star
  (or B star).  The thin distributions of early, mid, and late-type O
  stars (green, orange, and blue, respectively), as well as the dashed
  purple distribution of LBVs, are the same as in \citet{st15}.  These
  are distributions of separations on the sky to the nearest
  spectroscopically confirmed O star of any subtype or luminosity
  class. The thicker solid distributions, however, exclude from these
  samples all stars within 10 arcmin of the center of 30 Dor
  (following A19).  The dashed black distribution is for
  spectroscopically confirmed early B-type stars (ranging from later
  than O9.5 up to and including B2), measuring the separation to the
  nearest other B-type star in that same sample.  This ``early B''
  sample also excludes any stars with an apparent $V$ magnitude
  fainter than 13.9 mag (again following A19).  The solid black
  distribution is the same spectroscopic early B-star sample, but
  excluding all stars within 10 arcmin of the center of 30 Dor.  The
  magenta distribution is the BBS sample from A19, which is
  indistinguishable from the sample of spectroscopically confirmed
  early B stars.}
\label{fig:cumplot}
\end{figure*}

\section{EXCLUDING 30 DOR}

One potential source of bias in the photometric BBS arises because A19
made a choice to exclude all stars within a 10 arcmin radius of the
30~Dor region.  The reason for this choice was that they expected
crowding to be severe in 30 Dor, possibly compromising the
ground-based photometry. A19 did not evaluate the effect that this
exclusion might have on the resulting statistics.  There is cause for
potential concern, since this region around 30 Dor contains about half
of the known O-type stars in the LMC \citep{dekoter11} and most of the
known early O-type stars that are the putative progenitors of LBVs in
single-star models. Stars in the central regions of 30 Dor are among
the most densely clustered O-type stars, so excluding them might
selectively remove stars from the small end of the separation
distribution, shifting the median to larger separations.  On the other
hand, if crowding is severe and some of the most densely clustered
stars are missed, excluding 30~Dor might not have much impact, because
these stars are already undercounted.  This is straightforward to
test.

Figure~\ref{fig:cumplot} shows a cumulative distribution plot for
separations of O stars to the nearest other O-type star.  The thin
green, orange, and blue lines are early, mid, and late O-types stars,
respectively, which are essentially the same as in the original sample
of \citet{st15}.  The thicker lines of the same colors show what
happens to these distributions when we remove all the stars within a
10 arcmin radius from the center of 30 Dor.  The result is that the
O-type distributions are indeed skewed to larger separations as
qualitatively expected, but not by much.  The effect is more
significant for early O subtypes (a factor of $\sim$2 in separation
and implied age).  The exclusion of 30 Dor has less of an effect on
the separation distributions for mid and late O types, perhaps because
these samples of later O types are highly incomplete in the most
crowded regions.  Figure~\ref{fig:cumplot} also shows how this
exclusion influences the LBV separation distribution (thin dashed
purple vs. thick solid purple line), making the point that it has no
significant effect.  This is because most LBVs are not in clusters
anyway.

Thus, while excluding 30 Dor does skew the statistical distributions
to larger separations, it is not a large enough effect to fully
explain the discrepancy between spectroscopic O stars and the BBS
sample.  This is somewhat reassuring, as it indicates that despite the
large number of O-type stars in 30 Dor, there is nothing particularly
anomolous about the clustering {\it distribution} of O stars there,
and so it seems to be representative of O stars in general.  In other
words, outside 30 Dor in the rest of the LMC, O stars follow the same
pattern of being highly concentrated in clusters.  For early O-type
stars outside 30 Dor, the median separation is 5-6 pc and mid and late
O stars somewhat larger, still in good agreement with expectations
from models of cluster dispersal with age \citep{mojgan17}, and in
good agreement with general expectations for O stars residing in
clusters.

This exercise of excluding 30 Dor does highlight an interesting point
about LBVs, however, concerning total numbers.  Excluding 30~Dor
rejects about {\it half} the known O-type stars, and {\it most} of the
early O-type stars as noted above.  In stark contrast, excluding 30
Dor only removes 1 out of 26 LBVs in the sample (4\%). (That one LBV
is R143.)  The vast majority of LBVs (25/26) are not located in the
most active region of star formation in the LMC.  This underscores the
crucial point \citep{st15} that LBVs preferentially avoid O star
clusters.

If spectroscopically confirmed O stars are highly clustered as
expected both inside and outside 30 Dor, why, then, does the BBS
sample have such a large median separation of 31 pc?  Something else
is needed to reconcile the large difference between the median
separations of known O stars and the BBS sample.  As discussed below,
this is most likely because the photometrically selected BBS sample is
contaminated by an older population and does not trace the spatial
distribution of the most massive unevolved stars.

\begin{figure*}
\includegraphics[width=7.0in]{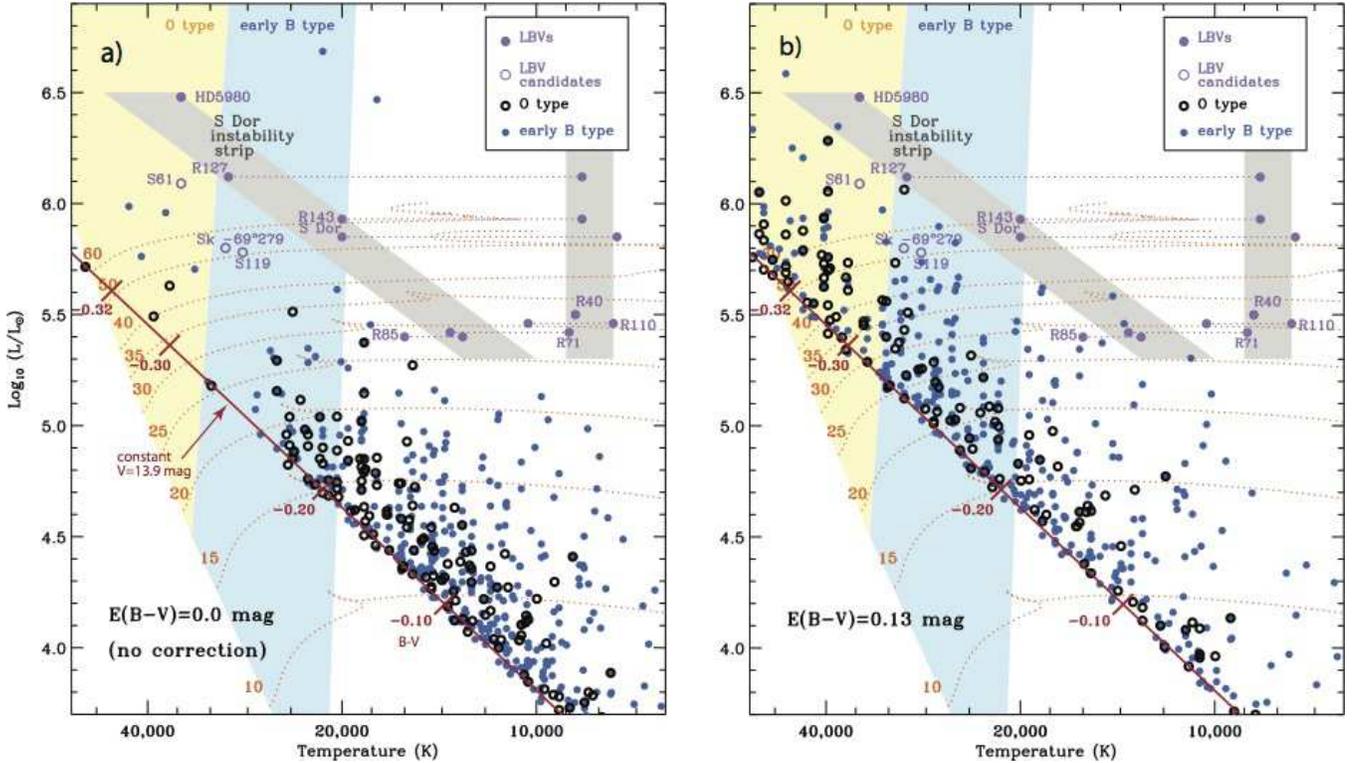}
\caption{HR Diagrams comparing LBVs to inferred properties of OB stars
  derived from photometric colors.  The $T_{\rm eff}$ value used to
  plot each OB star is the temperature one would infer by converting
  the apparent $B-V$ color to a temperature, and the luminosity comes
  from the bolometric correction for that $T_{\rm eff}$ value
  (relations adopted from \citealt{torres10} and \citealt{flower96})
  and its apparent $V$ magnitude.  The left panel (a) shows the result
  when no reddening correction is adopted, whereas the right panel (b)
  shows the values after applying a single average correction for
  $E(B-V)$=0.13 mag to all OB stars.  The black unfilled circles are
  spectroscopically confirmed O-type stars (similar to the sample in
  ST15), and the blue filled circles are spectroscopically confirmed
  early B-type stars, with spectral types between O9.5 and B2 (see
  text).  Both samples are limited to spectroscopically confirmed O or
  early B stars brighter than $V$ = 13.9 mag (to be consistent with
  the value adopted by A19), corresponding roughly to an absolute $V$
  magnitude of $M_V = -5$.  The diagonal red line in each panel shows
  the cutoff of $V$=13.9 mag, and the four red hash marks show
  temperatures corresponding to different values of $(B-V)$ = $-$0.32,
  $-$0.30, $-$0.20, and $-$0.10 mag.  In both panels, the orange
  dashed lines are single-star evolutionary tracks from
  \citet{brott11}, with initial masses in $M_{\odot}$ labeled in
  orange.  Light yellow and blue shaded areas indicate expected
  effective temperature and luminosity ranges for O-type and early
  B-type stars \citep{crowther06}, respectively.  LBVs with estimated
  $T_{\rm eff}$ and $L_{\rm Bol}$ values in the LMC and the S Doradus
  instability strip are included for reference, taken from
  \citet{smith19}.  The main point of this figure is to demonstrate
  that the resulting $T_{\rm eff}$ value that one infers from the
  apparent $B-V$ color has little to do with the star's true
  temperature; it is mainly determined by reddening (or lack thereof),
  since all these stars have small differences in their intrinsic
  $B-V$ color, but a relatively large spread in $E(B-V)$ from one
  object to the next.  Different values of reddening, whether
  corrected by a single average $E(B-V)$ value or with no correction
  (panels $b$ and $a$, respectively), lead to a huge spread in the
  inferred temperature that is much larger than the true temperature
  range of these spectroscopicaly selected O and early B-type stars
  (shaded yellow and blue areas). Importantly, when applying a single
  reddening correction to all, it is inevitable that many cooler stars
  (B stars) will be artifically shifted into the O star regime, and
  will therefore contaminate any color-selected sample of the bluest
  stars.  Note that the majority of stars that make the $V$ mag cut
  are close to that cutoff, and not in the region of $>$40 $M_{\odot}$
  stars.}
\label{fig:hrd}
\end{figure*}

\section{CONTAMINATION IN THE PHOTOMETRIC BBS SAMPLE}

\subsection{Likely sources of contamination} 

Concerned that the spectroscopic coverage of O-type stars in the LMC
might be spotty (past efforts to obtain spectra may have focussed on
clusters while neglecting field stars, for example), A19 aimed to
create a more complete sample of the most massive unevolved stars
using broad-brush photometric criteria.  However, ``more complete''
can also mean ``more contaminated'', because the youngest luminous
O-type stars and older luminous B-type supergiants have essentially
the same color at visual wavelengths.  While the broad brush technique
might be more inclusive of all the bright O-type stars, it may sweep
up many other blue stars that are not necessarily the most massive
unevolved stars.  As demonstrated below, the dangers of unavoidable
contamination in a photometric sample of blue stars outweighs the
beneft of higher completeness, and artificially skews the result. A19
adopted $UBV$ photometric criteria intended to provide a sample of the
most massive unevolved stars that would be complete for initial masses
above 40 $M_{\odot}$.  However, as emphasized by \citet{massey95}, it
is not possible to do this reliably (i.e. without contamination) using
photometry alone.  There are several compounding concerns, all of
which can push a photometric sample in the same direction of more
contamination from older, less massive stars:

{\it Degenerate colors:} The chief difficulty in selecting out the
hottest stars with photometry is degeneracy: O stars and early B-type
stars over a wide range of effective temperatures have essentially the
same intrinsic broad-band colors and magnitudes at visual wavelengths,
because $UBV$ photometry samples only the Rayleigh-Jeans tail of a hot
star's spectral energy distributon.  Their $B-V$ colors differ by only
about 0.02-0.03 mag from early O to early B types.  This is comparable
to or smaller than the photometric uncertainty in a single filter for
many stars in the data used by A19 (see below).  Spectra are needed to
reliably distinguish young and initially massive O-type stars apart
from older and less massive B supergiants.

{\it Binaries:} Compounding this problem of contamination is the fact
that massive stars are mostly in binary systems
\citep{blaauw61,sana12,moe17}.  Even without interaction, an
unresolved binary in ground-based photometry can be brighter than it
deserves for its age and spectral type simply because there are
multiple stars, but it may have essentially the same color as a hotter
and more massive star.  Hence, a less massive, older star with a
companion will masquerede as a more massive, younger star in a
photometric sample.

{\it Relative numbers and lifetimes:} While A19 acknowledged that
their color cuts may allow some B stars to enter the sample, they
presumed that the contamination would be minimal, and that because of
the $V$ magnitude cut ($V$=13.9 mag, corresponding to $M_V$=$-$5 mag
with an average extinction correction) most of the B stars in their
sample would be very luminous B supergiants that are the most massive
stars at the end of the main-sequence.  However, a concern is that the
B supergiants that correspond to a $>$60 $M_{\odot}$ star will be
extremely rare, but lower-mass B supergiants may be far more numerous.
Because the initial mass function favors lower masses, and because of
longer lifetimes at lower initial masses, B supergiants of 20-30
$M_{\odot}$ that just barely make the $V$ mag cut and color cut
(perhaps legitimately, or by photometric error, see below) can
outnumber the very rare 40-100 $M_{\odot}$ stars that fleetingly pass
through this cooler phase in single-star models.  (In other words, one
might expect that most of the stars in the box are near the lower
boundary of the box; this is tested and confirmed below.)  Using the
photometric criteria adopted by A19, there is no way to exclude these
$\sim$20 $M_{\odot}$ B supergiants.

{\it Expectations from single-star models vs. a real population with
  binaries:} A related complication has to do with expectations for
contamination by cooler stars guided by single-star models. A19
expected that any B-type star contamination should be small, since
they spend a very small fraction of their lifetime passing through the
cooler end of the main sequence.  A19 therefore assume that their
color-magnitude selection space should be dominated by the most
massive unevolved stars at hotter temperatures.  This expectation,
however, depends on the assumption that single-star models adequately
describe a real population of massive stars.  A long-standing problem
in massive star evoluton has been the large observed number of blue
supergiants \citep{fg90,evans07}, which is not satisfactorily
explained by the single-star models \citep{ekstrom12} that A19 used
for comparison.  Binary models can produce larger numbers of
long-lived blue supergiants at older ages as a result of mass transfer
and mergers
\citep{menon17,eldridge08,eldridge17,justham14,dvb13,farrell19,ppod92}.
Contamination should therefore be higher than expected from
considering only single-star models.

{\it Bleeding from photometric errors:} If the color cut worked
perfectly as intended, there would be only O-type stars in the BBS sample.
However, one must also consider possible bleeding due to photometric
errors, differences in reddening, or other effects.  In this way it
might be possible that even older and redder BSGs than the quoted
color cut could contaminate the BBS sample.  To select their BBS
sample from the $UBV$ photometry of \citet{zaritsky04}, A19 relied
upon the so-called reddening-free index $Q$, defined as $Q = (U-B) -
0.72 \times (B-V)$.  A19 chose to restrict the sample to $Q < -0.88$
mag, intending to select only stars hotter than about 35,000 K.  If
effective, this color cut would correspond to O dwarf spectral types
of O8.5 or earlier \citep{martins02}.  As noted above, the actual
differences in the intrinsic color between an early or mid O-type star
and an early B-star are very small (of order 0.02-0.03 mag in $B-V$,
or about 0.05 mag in $Q$).  A significant problem, however, is that
this intrinsic color difference is smaller than the photometric
uncertainty and the corresponding color uncertainty, making such a cut
unreliable for the goal of selecting only the most massive unevolved
stars.  \citet{zaritsky02,zaritsky04} quote zero point uncertainties
of 0.02 mag in $B$ and $V$, and 0.03-0.04 mag in $U$, and they note
typical rms scatters of $\sigma_U$=0.13, $\sigma_B$=0.07, and
$\sigma_V$=0.06 mag.  These typical uncertainties, even in a single
filter (and significantly worse in the resulting $Q$ value or $B-V$
color) are larger than the small color differences one is trying to
select against.  Morover, \citet{zaritsky04} note that stars in their
catalog that are brighter than 13.5 mag in $B$ or $V$ are prone to
``substantial photometric uncertainty'' (or flaring of 0.1 mag or
larger errors in various filters).  This substantial uncertainty
affects most of the stars in the BBS sample of A19, where the $V$ mag
cutoff was brighter than 13.9 mag.  This would seem to compromise the
ability to select the most massive unevolved stars by color.

Concerning this last point, A19 attempted to guard against bad
photometry by excluding stars with $Q < -$1.2 mag, being unphysically
blue, as well as excluding $U-B > -0.5$ mag, and $B-V > 0.2$ mag.
These criteria only prevent one from counting stars whose large errors
yield unrealistic colors (and they may also exclude very massive stars
that are highly reddened).  They do not, however, guard against cases
where a star's large photometric errors may shift it {\it inside} the
color-magnitude cut, even though its true temperature and luminosity
may belong outside.  The problem is that this type of contamination
can be severe because the very massive stars that are the intended
target of the photometric cuts are extremely rare, whereas they are
vastly outnumbered by lower-mass stars that should reside just outside
the cuts (see below).  As such, if even a small fraction of the stars
outside the cut can bleed in, they can strongly influence the median
age of the resulting sample.

\subsection{Expected contamination by B stars}

A19 conducted their analysis using the resulting photometric BBS
sample, noting that they ``expect them all to be high-mass stars,
primarily of O-type.''  Below we argue, however, that quantitative
tests of contamination were needed, because it is impossible to
photometrically select only the hottest stars when color differences
are small compared to photometric errors or reddening variation.
Moreover, some simple tests confirm that the BBS photometric sample
should be heavily contaminated by an older population, and therefore
unable to address the question of clustering and youth of LBVs.


Here we provide a brief illustration of the problem using O and B
stars with known spectral types.  We create a pseudo-BBS sample,
drawing from O and early B stars with known spectral types from
SIMBAD.\footnote{\tt http://simbad.u-strasbg.fr/simbad/} We use these
known spectral types as a rough indicator of the true stellar
temperature, to check against temperature inferred from photometric
colors.  As in \citet{st15}, we take all O-type stars in the LMC with
spectral types earlier than O9.5.  For the early B-types, we take
stars with spectral types between O9.5 and B2.  Both classes are
restricted to stars with apparent $V$ magnitudes brighter than 13.9,
as in A19, and we exclude stars within 10{\arcmin} of 30 Dor.  In this
resulting spectroscopically selected pseudo-BBS sample, the early
B-type stars outnumber the O-type stars roughly 3 to 1, although the
implications of this are unclear since we don't know the level of
incompleteness for either.

Figure~\ref{fig:hrd} shows an HR diagram with the spectroscopic OB
stars plotted without (left) and with (right) an average reddening
correction of $E(B-V)$=0.13 mag applied (the average value for OB
stars in the LMC adopted by A19).  Also shown for comparison are LBV
stars in the LMC and SMC, plotted using $T_{\rm eff}$ and $L_{\rm
  Bol}$ values from the literature \citep[see][]{smith19}, and
single-star evolutionary tracks \citep{brott11}.  OB stars are placed
on this HR diagram by taking their apparent or reddening corrected
$B-V$ color as a proxy for temperature, and the luminosity comes from
the $V$ mag and a bolometric correction for the corresponding $T_{\rm
  eff}$ value, with relations adopted from \citet{torres10} and
\citep{flower96}.  These are by no means intended to be taken as
accurate $T_{\rm eff}$ and $L_{\rm Bol}$ values; they are meant to
illustrate the range of these values one might infer from apparent
magnitudes and colors when only photometry is available.  The red
diagonal lines in Figure~\ref{fig:hrd} indicate where the $V$ mag
cutoff resides.  Even though these are not the same stars as in the
photometric BBS sample, there are several salient points that one can
glean from Figure~\ref{fig:hrd}.

First, the true temperature (indicated by the spectral type) has
little or nothing to do with the temperature inferred from photometric
colors. O and early B stars are thoroughly mixed with one another in
Figure~\ref{fig:hrd}, and it is impossible to differentiate hotter
O-type stars from cooler early B-type stars based on color.  The
resulting range of temperatures is entirely a result of different
reddening along individual lines of sight, plus photometric error.
Adopting a single average value for the reddening is clearly invalid,
and applying such a reddening correction simply shifts both swarms of
O and B stars to higher inferred temperatures and luminosities.  One
would infer from Figure~\ref{fig:hrd} that a color selection of the
bluest stars would result in a sample that is split between O and
early B stars (this is indeed the case for the BBS sample, see below),
if not dominated by early B-type stars.

Second, Figure~\ref{fig:hrd} confirms that the $V$ mag cutoff does not
reliably select the most massive stars.  Among OB stars that make the
$V$ mag cut, many are concentrated near the faint cutoff.  If we were
to deredden the early B-type stars to their appropriate temperatures
of 20-30 kK, they would mostly land along evolutionary tracks for
20-30 $M_{\odot}$ stars, not $>$40 $M_{\odot}$ stars.  Without
spectra, there is no way to reliably estimate the star's temperature
and luminosity at the precision needed to distinguish a cooler 20-30
$M_{\odot}$ B supergiant star from a hotter and more massive O-type
star.  Using visual-wavelength color-magnitude cuts to produce a
sample of the most massive unevolved stars is therefore invalid for
the purpose of testing LBV evolution.

Figure~\ref{fig:hrd} also presents a cautionary tale against using
apparent $B-V$ colors alone as a proxy for temperature.\footnote{A19
  used a $Q$ parameter selection as noted above, so the rest of this
  paragraph does not apply specifically to that study.  However, some
  studies of stellar populations use only $B-V$ or $V-I$, for example,
  to infer properties about the stellar population.}  While all O and
early B stars have nearly identical intrinsic $B-V$ colors, much
larger differences in reddening from one star to the next of only 0.1
mag or more in $E(B-V)$ can lead to a gigantic spread in the inferred
temperature that crosses much of HR diagram in Figure~\ref{fig:hrd}.
OB associations in the LMC have a wide range of different reddening
values, even varying significantly among individual OB stars in the
same association \citep{lucke74}.  Selecting by blue color and bright
$V$ mag will favor those with the least reddening, not necessarily the
hottest, youngest, or most luminous stars.  This could potentially
yield a systematic bias against the highest-mass stars, and to instead
preferentially select older, lower-mass, evolved blue supergiants in a
population.  This is because evolved BSGs with longer lifetimes are
more likely to have cleared away or drifted away from their
surrounding natal clouds, and may therefore have bluer apparent colors
because of lower reddening.  The youngest and most massive stars,
which have essentially the same intrinsic $B-V$ color, are more likely
to still be partly embedded, and therefore more reddened by dust from
their surrounding natal environment that has not yet cleared away
\citep{bw59,reddish67,ys01}.

\subsection{Confirmed contamination by B stars}

So far this evaluation has been hypothetical; i.e. considerations for
why there {\it could be} or {\it should be} contamination of a
photometrically selected sample of bright blue stars.  In fact, it has
already been demonstrated observationally that the photometric BBS
sample is strongly contaminated by cooler B supergiant stars.  A19
noted that among their BBS sample, about half the stars have known
spectral types available, while the other half have no spectral types.
Of those BBS stars with available spectra, slightly less than half are
O stars (135 stars, or 49\%) and slightly more than half are early B
stars (140 stars, or 51\%).  So although the intent of the color
selection was to include only massive stars hotter than $T_{\rm eff}$
= 35 kK (corresponding to dwarfs of type O8.5 and earlier, as noted
above), the BBS sample nevertheless includes many cooler stars
(early
B supergiants have $T_{\rm eff}$ values of roughly 20 kK to 30 kK;
\citealt{crowther06,crowther08}).

For B-type stars included in the BBS sample, A19 assumed that these
would be limited to the most massive stars near the end of the main
sequence due to the bright $V$ mag cutoff.  However, examining
Figure~\ref{fig:hrd}, it is clear that most of the early B stars that
satisfy the $V$ mag cut and would make it into the sample, by number,
are not the most massive stars that are at the end of the main
sequence.  Instead, most are stars clustered near the $V$ mag cutoff
that just barely pass the cut.  These are overwhelmingly lower-mass
evolved BSGs. This contamination dramatically alters the resulting
distribution of separations.  Overlooking this contamination
undermines the analysis, as demonstrated below.

Moreover, even for those spectroscopically confirmed O stars in the
BBS sample, a majority might be later O-types (O8, O9), which make up
the majority of O stars by number.  This distinction is important,
since later O type stars can have much longer lifetimes and may come
from lower initial masses than early O-types, as seen from the
breakdown of separations for late vs. early O subtypes \citep{st15}.
Similarly, most of the spectroscopically confirmed O-type stars that
satisfy the $V$ mag cut in Figure~\ref{fig:hrd} are at the
low-luminosity boundary near the $V$ mag cutoff, not in the region
corresponding to $>$60 $M_{\odot}$ stars.

Interestingly, examining Table~1 from \citet{st15}, the nearest or
second nearest spectroscopically selected O-type star to each LBV is,
in the vast majority of cases, a late-type O star (O8/O9) and not an
early O-type star.  Since the median separation of LBVs is the same
for the spectroscopic O sample \citep{st15} and the photometric BBS
sample (A19), we can surmise that these nearest neighbors are in many
cases the same stars.

\section{CONTAMINATION EXPLAINS THE BBS SEPARATION DISTRIBUTION}

From the discussion above, it is clear that the BBS sample is
contaminated by older, evolved B supergiant stars, rather than being
restricted to only the most massive unevolved main-sequence stars that
the sample was intended to trace.  The next question to ask is whether
such contamination could plausibly explain the observed large
separation distribution of BBS stars that A19 found.  One can address
this by asking the pertinent question: What does the distribution of
projected separations on the sky look like for stars that we know are
cooler, evolved B-type supergiants?  How does it compare to the BBS
sample and to O-type stars?

One can test this by considering a sample of stars that are known to
be early B supergiants because they are confirmed by spectroscopy.  A
sample of spectroscopically confirmed early B supergiant stars in the
LMC was extracted from SIMBAD, as noted above, the same way that
\citet{st15} produced spectroscopically confirmed samples of O-type
stars.  We chose this ``early B'' sample to include LMC stars with
spectral types later than O9.5 and up to B2, of any luminosity class.
This range was chosen because their small differences in intrinsic
color compared to O-type stars would pass the selection criteria of
A19, especially considering photometric errors.  The spectroscopic
sample was also restricted to an apparent $V$ magnitude cutoff
brighter than 13.9 mag, to be consistent with the BBS sample of A19,
and therefore selects primarily B supergiants.  With $T_{\rm eff}$
values as low as 20 kK, this sample will include a large number of
evolved B supergiants with initial masses around 20 $M_{\odot}$, and
possibly even down to 15 $M_{\odot}$.  This sample of early B
supergiants is plotted alongside spectroscopically confirmed O-type
stars in Figure~\ref{fig:hrd}.  Importantly, one can see from
Figure~\ref{fig:hrd} that only a handful of these B supergiants are
luminous enough to be late-main sequence stars of $>$40 $M_{\odot}$;
the vast majority of the early B supergiants that make the $V$ mag cut
are consistent with less massive stars (15-30 $M_{\odot}$ single-star
tracks) and are therefore older than presumed single-star progenitors
of LBVs.

The resulting cumulative distribution of separations for these early B
stars is shown in Figure~\ref{fig:cumplot}.  The black cumulative
distribution in Figure~\ref{fig:cumplot} is for this sample of
spectroscopically confirmed early B-type stars, where the separation
is measured to the nearest other early B star in the same sample (not
the nearest O star).  Note that, pertinent to the discussion in
Section 3 above, the {\it dashed} black distribution is for all the
known early B stars in this sample, whereas the {\it thick solid}
black distribution excludes stars within 10{\arcmin} of 30 Dor (to be
consistent with A19).  There is little difference, because B
supergiants are not clustered on small scales.

The most interesting result here is that the separation distribution
of this sample of spectroscopically confirmed early B supergiants
matches that of the photometric BBS sample (shown in magenta in
Figure~\ref{fig:cumplot}) from A19.  One may debate if the potential
sources of bias and contamination discussed above are actually to
blame, or if some other effects are important. But whatever the exact
reason, the outcome confirms that the photometric BBS separation
distribution is characteristic of an older population than expected
for LBVs, because its median separation is identical to stars that are
spectroscopically confirmed to be cooler, evolved, lower-mass stars.
The BBS separation distribution is clearly incompatible with
confirmed early O-type stars that are known to be the most massive
unevolved stars.\footnote{Note that the resulting large separation
  distribution of spectroscopically confirmed early B-type stars in
  Figure~\ref{fig:cumplot} would seem to contradict the presumption
  that past efforts to obtain spectral types have been heavily biased
  toward clustered regions.  It was already demonstrated in Section 2
  that any such incompleteness in the spectroscopic sample does not
  impact the result.}  Thus, one may conclude that contamination by
older stars is the dominant explanation for the large median
separation of the BBS sample and its consequent similarity to LBVs.
In other words, the similarity between the BBS and LBV separation
distributions arises because the photometric BBS sample is old, not
because the LBVs are young.

Although harder to demonstrate in the same way as above for the LMC,
it is quite likely that the same conclusion about the age of bright
blue stars applies to the BBS samples for M31 and M33 that A19
discussed.  For BBS stars in M31/M33, A19 found a median separation of
65 pc, indicating that this BBS sample clearly does not trace a
clustered reference population either.  The M31/M33 sample from
ground-based photometry may also have the added drawback of inadequate
angular resolution to trace clustered young stars.  Since it is not
tracing clustered stars, it fails the requirement to use spatial
dispersal on the sky as a relative age indicator.  Those M31/M33
distributions are therefore similarly not indicative of LBV youth.

\section{WR and WN3/O3 STARS}

As noted above in Section 2, one of the key requirements for the
separation distribution method to work as an age indicator is that the
reference population {\it must} be tracing a clustered population.  If
the reference sample is not clustered, then the resulting separation
distribution is simply not measuring a relative age.  Consider the
alternative in the following {\it gedanken}-experiment: Imagine an
evenly-spaced grid of blue stars that is distributed over a portion of
the sky with adjacent stars each separated by $\sim$30 pc.  Now
randomly drop in a less numerous population of stars (either WR stars
or LBVs, for example) and measure the resulting separation
distribution.  One will find that the separation between WR stars and
the nearest blue stars, or the separation between LBVs and the nearest
blue star, will both also tend to be around 30 pc.  This is not
providing information about the relative clustering or youth of the WR
stars or the LBVs; it is merely indicating the typical separation
between one blue reference star and the next (i.e. one cannot find a
median separation from a WR star or LBV to a blue star much different
from 30 pc, because that is the grid spacing).  With a dispersed
sample serving as the comparison, distributions get squeezed together
on a separation plot because the test is not precise enough to
distinguish differences in age.

This explains why A19 found BBS stars, LBVs, and WR stars to all have
roughly the same separation distribution (see their Figure 2).  A19
noted that a KS test showed no statistical difference between them.
Rather than indicating that all three groups are young and consistent
with the evolution of the most massive single stars, this similarity
is merely tracing the typical separation between BBS stars themselves.
If a clear difference in separation distribution is not revealed by
this comparison, then it is incorrect to conclude that the samples are
all equally young --- one may only conclude that the test is not
precise enough.  A19 did find a significant difference in the
resulting separation distribution of red supergiants (RSGs), but this
is probably because RSGs with typical initial masses of only 9
$M_{\odot}$ in their sample are much older than the BBS stars, and
tend to occupy regions of the LMC where most blue supergiants are
long-since dead.

This issue of how weak clustering in the reference sample will
undermine the outcome also resolves a recent debate in the literature
about WN3/O3 stars based on their separation
distribution on the sky.  WN3/O3 stars are a subclass of WR stars
found in the LMC, which, like typical WR stars in the SMC, have
transitional spectra with both emission and absorption lines
\citep{massey14}.

Using the same methodology that \citet{st15} used for LBVs,
\citet{smith18} examined the distribution of separations between
WN3/O3 stars and spectroscopically confirmed O-type stars in the LMC.
They found that WN3/O3 stars are extremely isolated from clustered O
stars (even more so than LBVs), having a distribution on the sky
similar to 15-20 $M_{\odot}$ RSGs.  This makes it unlikely that WN3/O3
stars are very massive stars that have evolved as rapidly rotating
single stars through quasi-chemically homogeneous evolution or wind
mass loss.  Instead, \citet{smith18} proposed that they arise from
moderately massive (15-20 $M_{\odot}$) progenitors that have had their
H envelopes stripped through interaction with a lower-mass companion
star.  \citet{gotberg18} demonstrated that such stars arise naturally
in a grid of binary evolution models with model atmospheres. They
occur in a transitional zone at the low-mass and low-luminosity end of
the range of normal WR stars that form in binaries, where winds are
still dense enough to have emission lines, but are thin enough to also
see absorption lines in the underlying hot photosphere.  At LMC
metallicity, these stars are expected to arise from initial masses
around 15-20 $M_{\odot}$ \citep{gotberg18}, in good agreement with the
observed spatial distribution of WN3/O3 stars \citep{smith18}.  As
such, the WN3/O3 stars would be of interest as candidates for common
progenitors of stripped-envelope SNe.

A debate that echoes the one over LBV separation and ages also arose
for these WN3/O3 stars.  \citet{neugent18} re-examined the separation
distribution of WN3/O3 stars by comparing them to a photometrically
selected sample of bright blue stars, and much like A19 with LBVs,
found them to be less isolated from these stars than when they are
compared to spectroscopically confirmed O-type stars \citep{smith18}.
\citet{neugent18} also
chose a photometrically selected comparison sample of bright blue
stars from the same \citet{zaritsky04} photometric catalog; their selection
criteria were similar to the criteria adopted by A19, although
somewhat more relaxed (with $V < 15$ mag instead of $V < 13.9$ mag,
and less restrictive colors with $Q < -0.80$ mag instead of $Q <
-0.88$ mag, for example).

One might anticipate that the photometric comparison sample of blue
stars used by \citet{neugent18} falls victim to the same pitfalls as
the BBS sample used by A19, for the same reasons discussed above.
First, it will be contaminated by less massive B supergiants with
initial masses around 15-20 $M_{\odot}$, similar to the proposed
binary initial masses of WN3/O3 stars \citep{smith18}. The
contamination by lower-mass B supergiants is likely to be even more
severe than the BBS sample of A19, because the BBS sample from
\citet{neugent18} goes a magnitude fainter in $V$ and accepts redder
stars.  Second, this photometric sample of blue stars is also not
clustered, and therefore cannot be used to test relative ages.
\citet{neugent18} found that their comparison to photometric blue star
locations yielded separations for WN3/O3 stars (156{\arcsec} or 37 pc)
that were statistically indistinguishable in a KS test from those of
classical early WN stars (110{\arcsec} or 26 pc).  These are both
similar to the median separation between BBS stars and other BBS stars
of 31 pc (A19).
Following the same explanation as discussed above for the BBS sample
of A19, the likely reason that these distributions are all so similar
is because the measured separation distributions of WR stars are
limited by the typical separation amongst the unclustered blue stars
themselves.

\section{SUMMARY AND DISCUSSION}

Using the dispersal on the sky can be a powerful way to rank relative
ages of different classes of stars and to discriminate between single
and binary evolution channels, but this particular method only works
as an age indicator if one is comparing to a reference population that
(1) is known to be young and (2) is highly clustered.  If both of
these criteria are not clearly met, then the results of the comparison
are invalid.  The method relies upon the assumption that massive stars
begin their lives mostly in clusters, and that their relative
separation increases with time because clusters disperse and the most
massive stars die quickly.

This method was initially used to demonstrate that LBVs are more
isolated than they should be in the traditional scenario where they
have massive single-star progenitors.  The proposed reason was either
because they have received kicks from a companion's SN, or because
they have been rejuvenated by mass transfer or mergers in binaries
\citep{st15,smith16,mojgan17}.  Either case requires that LBVs are
mainly a product of close binary evolution.  This was inferred using
spectroscopically confirmed O stars as a tracer of clustered young
stars, revealing that LBVs clearly do not reside in the same places in
the sky as known O-type stars.

A19 recently conducted a similar type of study, but drew the opposite
conclusion for LBVs -- i.e. that their distribution on the sky does
not contradict a single-star evolutionary scenario.  This was because
A19 found the separation distribution of LBVs to be not too different
from that of a photometric sample of bright blue stars (BBS).
Arriving at this interpretation, A19 assumed that the photometric BBS
sample was reliably tracing the most massive unevolved stars, so that
the similar separation distribution of LBVs would imply that LBVs are
not so old.  On the other hand, if the BBS sample was tracing older
stars, then one draws the opposite conclusion.

The preceding sections of this paper have discussed ways in which the
single-star interpretation is problematic, largely because the two key
critieria for this type of study to work (outlined above in Section 2)
are not met by the photometric BBS sample:

(1) First, the stars in the BBS sample are not confidently known to be
young, because $UBV$ photometric color selection does not provide a
robust way to separate the youngest, hottest, and most massive O stars
from older, somewhat evolved, less massive B and late-O supergiants.
This is because they all have similar intrinsic color. Differences in
intrinsic color are less than effects of reddening and photometric
errors.  Being able to make this distinction reliably is, however,
critical to the interpretation, because it links to the difference
between single and binary progenitor scenarios for LBVs.  If LBVs
descend from massive single stars, they should have ages similar to
the most massive unevolved early O-type stars.  If LBVs descend from
lower mass binaries that experience rejuvenation though mass transfer
or mergers, then they should have a true age that is similar to the
lower-mass B and late-O supergiants.  If LBVs have a separation
distribution that is similar to a photometric BBS sample, then one's
conclusion can flip depending on whether or not the BBS sample is
really dominated by the youngest most massive stars.
In fact, it is clear that the BBS sample is contaminated at a
substantial level, because among the half of the BBS sample with
available spectra, more than half of these are B supergiants (many of
the remainder are probably late-O supergiants).  This confirms that
substantial contamination by older stars has occurred, despite the
intent of restrictive photometric selection.  This is at least partly
due to the fact that the photometric errors were larger than intrinsic
color differences.

(2) Second, the stars in the BBS sample are not highly clustered, and
so they cannot be used as a reliable reference for diagnosing youth.
The median separation from a BBS star to its nearest neighbor in the
LMC is 31 pc (or 65 pc for BBS stars in M31/M33), meaning that the BBS
sample is not concentrated in young dense clusters, as O stars are
known to be.  Less than 4\% of the BBS sample has a separation
commensurate with being in young massive clusters, whereas the
spectroscopic O-star sample matches expectations for young O stars.
Instead, the distribution of BBS stars is matched well by the observed
separation distribution of spectroscopically confirmed early B
supergiants.  This gives a strong confirmation that whatever the
intent, the photometric BBS sample ends up tracing the separation
distribution of an older population of evolved, lower-mass stars, and
not the presumed massive single-star progenitors of LBVs.  Invoking
incompleteness of the spectroscopic O-star sample does not explain why
the BBS sample is so unclustered.

Overall, LBV environments match quite well expectations for binary
evolution, where LBVs are massive blue stragglers produced either by
mass accretion from companion and possible kicks, or by rejuvenation
in stellar mergers \citep{st15,smith16,mojgan17}.  If one is willing
to accept the notion that close binaries are so common that they may
dominate the evolutionary paths of massive stars
\citep{sana12,demink14,moe17,eldridge17}, then this result is not so
surprising.  Quantitatively, the median separation of the BBS sample
matches expectations from a simple dispersing cluster model for ages
of 9-10 Myr \citep{mojgan17}, and it matches the observed separation
distribution of known B-type supergiants, as noted above.  LBVs have a
similar separation distribution indicating that they are this old as
well, or older.  This could not be the case if LBVs occur immediately
after the main sequence in massive stars with lifetimes of only 3-4
Myr.

More broadly, the analysis above offers a cautionary tale when
analyzing photometric samples of bright blue stars in resolved stellar
populations.  The typical bright blue star is more likely to be an
evolved, moderately massive blue supergiant (akin to the $\sim$18
$M_{\odot}$ progenitor of SN~1987A; \citealt{arnett89}) and is less
likely to be a very young, very massive main-sequence O-type star.
Both types of stars have almost the same instrinsic colors and
magnitudes at visual wavelengths, but those with earlier spectral
types, higher luminosity, and higher initial mass are disfavored by
number due to the initial mass function and shorter lifetimes.
Similarly, in more distant galaxies with unresolved stellar
populations, this implies that a blue color in surrounding galaxy
light will tend to favor an age of 10-15 Myr, rather than a very young
population around 3-4 Myr.  Bluer does not necessarily mean younger.
This is important for interpreting the relative ages and initial
masses we associate with different SN types based on their surrounding
host color, for example \citep{kk12}.  As has been suggested for the
progenitor of SN~1987A \citep{ppod10}, many of these BSGs that produce
the blue light in stellar populations may be products of binary
interaction (mergers and mass gainers, producing blue stragglers), and
they may therefore be more common than one might expect from a
single-star population.  Anecdotally, it is interesting to note that
the most distant multiply lensed SN that has been detected was a
SN~1987A-like event from a BSG \citep{kelly16}, and the most distant
individual lensed star (in that same host galaxy) appears consistent
with a BSG \citep{kelly18}.

\section*{Acknowledgements}

\scriptsize I thank an anonymous referee for a careful reading of the
manuscript and suggestions that improved the paper.  I benefitted from
useful discussions with Mojgan Aghakhanloo, Ylva G\"otberg, Jeremiah
Murphy, Selma de Mink, Maria Drout, Pat Kelly, Matteo Cantiello, Armin
Rest, and Ofer Yaron.  Support for NS was provided by NSF awards
AST-1312221 and AST-1515559, and by the National Aeronautics and Space
Administration (NASA) through HST grant AR-14316 from the Space
Telescope Science Institute, which is operated by AURA, Inc., under
NASA contract NAS5-26555.  This research has made use of the SIMBAD
data base, operated at CDS, Strasbourg, France.


\begin{thebibliography}{}

\bibitem[Aadland et al.(2018)]{aadland18} Aadland E, Massey P, Neugent
  KF, Drout MR.  2018, AJ, 156, 294 (A19)
  
\bibitem[Aghakhanloo et al.(2017)]{mojgan17}
  Aghakhanloo M, Murphy J, Smith Nm Hlozek R. 2017, MNRAS,
  472, 591



\bibitem[Arnett et al.(1989)]{arnett89} Arnett WD, Bahcall JN,
  Kirshner RP, Woosley SE. 1989, ARA\&A, 27, 629

\bibitem[Beasor et al.(2019)]{bds19} Beasor, ER, Davies, B,
  Smith N, Bastian, N. 2019, MNRAS, 486, 266
  
\bibitem[Blaauw(1961)]{blaauw61} Blaauw A. 1961, BAN, 15, 265

\bibitem[Blaauw(1964)]{blaauw64} Blaauw A. 1964, ARAA, 2, 213
 
\bibitem[Blagovest et al.(2016)]{blagovest16} Blagovest P, Vink JS,
  Gr\"afener G. 2016, MNRAS, 458, 1999

\bibitem[Blanco \& Williams(1959)]{bw59} Blanco VM, Williams AD. 1959,
  ApJ, 130, 482
  

\bibitem[Bouret et al.(2005)]{bouret05} Bouret JC, Lanz T, Hillier
  DJ. 2005, A\&A, 438, 301

\bibitem[Brott et al.(2011)]{brott11} Brott, I., Evans, C. J., Hunter,
  I., et al. 2011, A\&A, 530, A115

\bibitem[Conti(1976)]{conti76} Conti PS. 1976, Mem.\ Soc.\ R.\ Sci.\
  Lie'ge, 9, 193


\bibitem[Crowther et al.(2006)]{crowther06} Crowther PA, Lennon DA,
  Walborn NR. 2006, A\&A, 446, 279


\bibitem[Crowther et al.(2008)]{crowther08} Crowther PA, Lennon DA,
  Walborn NR, Smartt SJ. 2008, Asp Conf Ser, 388, 109


\bibitem[de Koter et al.(2011)]{dekoter11} de Koter A, et al.\ 2011,
  J. Phys., Conf.\ Ser., 318, 012022
  
\bibitem[de Mink et al.(2014)]{demink14} de Mink SE, et al.\ 2014,
  ApJ, 782, 7


\bibitem[Ekstr\"om et al.(2012)]{ekstrom12} Ekstr\''om, S., Georgy,
  C., Eggenberger, P., et al. 2012, A\&A, 537, A146

\bibitem[Eldridge et al.(2008)]{eldridge08} Eldridge JJ, Izzard RG,
  Tout, CA. 2008, MNRAS, 384, 1109
  
\bibitem[Eldridge et al.(2011)]{eldridge11} Eldridge JJ, Langer N,
  Tout CA. 2011, MNRAS, 414, 3501

\bibitem[Eldridge et al.(2017)]{eldridge17} Eldridge JJ, Stanway ER,
  Xiao L, et al. 2017, PASA, 34, 58
  
\bibitem[Evans et al.(2007)]{evans07} Evans CJ, Lennon DJ, Dufton PL,
  Trundle C. 2007, A\&A, 464, 289
  
\bibitem[Farrell et a.(2019)]{farrell19} Farrell EJ, Groh JH, Meynet
  G, Kudritzki RP, Eldridge JJ, Georgy C, Ekstrom S, Yoon SC. 2019,
  A\&A, 621, A22

\bibitem[Fitzpatrick \& Garmany(1990)]{fg90} Fitzpatrick EL, Garmany
  CD. 1990, ApJ, 363, 119
  
\bibitem[Flower(1996)]{flower96} Flower PJ. 1996, ApJ, 469, 355
  
\bibitem[Fullerton et al.(2006)]{fullerton06} Fullerton AW, Massa DL,
  Prinja RK. 2006, ApJ, 637, 1025

\bibitem[Gallagher(1989)]{jsg89} Gallagher, J.S.\ 1989, in Physics of
  Luminous Blue Variables, ed. K.\ Davidson, A.F.J.\ Moffat, \&
  H.J.G.L.M.\ Lamers (Dordrect: Kluwer), 185



\bibitem[Garmany et al.(1982)]{garmany82} Garmany CD, Conti PS, Chiosi
  C. 1982, ApJ, 263, 777

\bibitem[Gies(1987)]{gies87} Gies DR. 1987, ApJS, 64, 545

\bibitem[G\"otberg et al.(2017)]{gotberg17} G\"otberg Y, de Mink SE,
  Groh JS. 2017, A\&A, 608, 11

\bibitem[G\"otberg et al.(2018)]{gotberg18} G\"otberg Y, de Mink SE,
  Groh JS, Kupfer T, Crowther PA, Zapartas E, Renzo M.  2018, A\&A,
  615, A78



  
  


\bibitem[Groh et al.(2013a)]{groh13} Groh JS, Meynet G, Georgy C,
  Ekstr\"{o}m S. 2013a, A\&A, 558, A131

\bibitem[Groh et al.(2013b)]{groh13b} Groh JS, Meynet G, Ekstr\"{o}m
  S. 2013b, A\&A, 550, L7

\bibitem[Harris \& Zaritsky(2009)]{hz09} Harris J, Zaritsky D. 2009,
  AJ, 138, 1243
   
\bibitem[Heger et al.(2003)]{heger03} Heger A., Fryer C. L., Woosley
  S. E., Langer N., Hartmann D. H., 2003, ApJ, 591, 288
  
\bibitem[Hubble \& Sandage(1953)]{hs53} Hubble E, Sandage A.\ 1953,
  ApJ, 118, 353

\bibitem[Humphreys \& Davidson(1994)]{hd94} Humphreys RM, Davidson K.\
  1994, PASP, 106, 1025


\bibitem[Humphreys et al.(2016)]{h16} Humphreys RM, Weis K, Davidson
  K, GordonMS. 2016, ApJ, 825, 64

\bibitem[Justham et al.(2014)]{justham14} Justham S, Podsiadlowski P,
  Vink JS. 2014, ApJ, 796, 121

\bibitem[Kelly \& Kirshner(2012)]{kk12} Kelly PL, Kirshner RP. 2012,
  ApJ, 759, 107


\bibitem[Kelly et al.(2016)]{kelly16} Kelly et al. 2016, ApJ, 831, 205

\bibitem[Kelly et al.(2018)]{kelly18} Kelly et al. 2018, Nature
  Astronomy, 2, 334


\bibitem[Kenyon \& Gallagher(1985)]{kg85} Kenyon, S.J., \& Gallagher,
  J.S.\ 1989, ApJ, 290, 542



\bibitem[King(2000)]{king00} King NL. 2000, Ph. D.Thesis, NMSU%

  
\bibitem[Langer et al.(1994)]{langer94} Langer N, Hamann WR, Lennon
  MD, Najarro F, Pauldrach AWA, Puls J. 1994, A\&A, 290, 819

\bibitem[Lortet(1989)]{lortet89} Lortet MC. 1989, in Physics of
  Luminous Blue Variables, ed. K.\ Davidson, A.F.J.\ Moffat, \&
  H.J.G.L.M.\ Lamers (Dordrect: Kluwer), 45

\bibitem[Lucke(1974)]{lucke74} Lucke PB. 1974, ApJS, 255, 28

\bibitem[Lynds(1980)]{lynds80} Lynds, B.T. 1980, AJ, 85, 1046

\bibitem[Maeder(1989)]{maeder89} Maeder A. 1989, in Physics of
  Luminous Blue Variables, ed. K.\ Davidson, A.F.J.\ Moffat, \&
  H.J.G.L.M.\ Lamers (Dordrect: Kluwer), 15

\bibitem[Martins et al.(2002)]{martins02} Martins F, Schaerer D,
  Hillier DJ. 2002, A\&A, 382, 999

\bibitem[Massey et al.(1995)]{massey95} Massey P, Lang CC,
  Degioia-Eastwood K, Garmnany CD. 1995, ApJ, 438, 188


\bibitem[Massey et al.(2007)]{massey07} Massey P., et al.\ 2007, AJ,
  134, 2474

\bibitem[Massey et al.(2014)]{massey14} Massey P, Neugent KF, Morrell
  N, Hillier DJ, 2014, ApJ, 788, 83

\bibitem[Menon \& Heger(2017)]{menon17} Menon A, Heger A. 2017, MNRAS,
  469, 4649
  
\bibitem[Meynet \& Maeder(2003)]{mm03} Meynet G, Maeder A. 2003, A\&A,
  404,975

\bibitem[Meynet et al.(2011)]{meynet11} Meynet G, Georgy C, Hirschi R,
  Maeder A, Massey P, Przybilla N, Nieva MF. 2011, Bull.\ Soc.\ R.\
  Sci.\ Lie'ge, 80, 266

\bibitem[Moe \& Di Stefano(2017)]{moe17} Moe M., Di Stefano R., 2017,
  ApJS, 230, 15
  
\bibitem[Neugent et al.(2018)]{neugent18} Neugent KF, Massey P,
  Morrell N. 2018, ApJ, 863, 181






  %

\bibitem[Owocki et al.(2004)]{owocki04} Owocki SP, et al. 2004, ApJ,
  616, 525

  
\bibitem[Paczynski(1961)]{paczynski61} Paczynski B. 1961, Acta
  Astron., 17, 355

\bibitem[Podsiadlowski(2010)]{ppod10} Podsiadlowski P. 2010, New
  Astron. Rev., 54, 39

\bibitem[Podsiadlowski et al.(1992)]{ppod92} Podsiadlowski P, Joss PC,
  Hsu JJL. 1992, ApJ, 391, 246
  

\bibitem[Reddish(1967)]{reddish67} Reddish VC. 1967, MNRAS, 135, 251

\bibitem[Renzo et al.(2019)]{renzo19} Renzo M, Zapartas E, de Mink SE,
  et al. 2019, A\&A, 624, A66
  
\bibitem[Sana et al.(2012)]{sana12} Sana H, de Mink SE, de Koter A, et
  al. 2012, Science, 337, 444

\bibitem[Schneider, Izzard, de Mink, et al.(2014)]{schneider14}
  Schneider FRN, Izzard RG, de Mink SE, et al. 2014, ApJ, 780, 117
  
  

\bibitem[Smith(2014)]{smith14} Smith N. 2014, ARAA, 52, 487

\bibitem[Smith(2016)]{smith16} Smith N. 2016, MNRAS, 461, 3353

\bibitem[Smith \& Arnett(2014)]{sa14} Smith N, Arnett D., 2014, ApJ,
  785, 82

  
\bibitem[Smith \& Conti(2008)]{sc08} Smith N, Conti PS. 2008, ApJ,
  679, 1467


\bibitem[Smith \& Owocki(2006)]{so06} Smith N, Owocki SP. 2006, ApJ,
  645, L45

\bibitem[Smith \& Tombleson(2015)]{st15} Smith N, Tombleson R. 2015,
  MNRAS, 447, 602






\bibitem[Smith et al.(2011)]{smith11} Smith N, Li W, Silverman JM,
  Ganeshalingam M, Filippenko AV.  2011a, MNRAS, 415, 773 



\bibitem[Smith et al.(2018)]{smith18} Smith N, G\"otberg Y, de Mink
  SE. 2018, MNRAS, 475, 772

\bibitem[Smith et al.(2019)]{smith19} Smith N, et al.\ 2019, MNRAS, in
  press (arXiv:1805.03298) 
  

 
\bibitem[Tammann \& Sandage(1968)]{ts68} Tammann GA, \& Sandage A
  1968, APJ, 151, 825

\bibitem[Torres(2010)]{torres10} Torres G. 2010, AJ, q40, 1158
  



\bibitem[Vanbeveren et al.(2013)]{dvb13} Vanbeveren D, Mennekens N,
  Van Rensbergen W, De Loore C. 2013, A\&A, 552, A105
  


\bibitem[van Genderen(2001)]{vg01}van Genderen AM. 2001, A\&A, 366,
  508



  


\bibitem[Yadav \& Sagar(2001)]{ys01} Yadav RKS, Sagar R. 2001, MNRAS,
  328, 370

\bibitem[Zaritsky et al.(2002)]{zaritsky02} Zaritsky D, Harris J,
  Thompson IB, Grebel EK, Massey P. 2002< AJ, 123, 855
  
\bibitem[Zaritsky et al.(2004)]{zaritsky04} Zaritsky D, Harris J,
  Thompson IB, Grebel EK. 2004, AJ, 128, 1606

\end{thebibliography}
\end{document}